\documentclass[runningheads]{svmult}
\usepackage{makeidx}   
\usepackage{graphicx}  
\usepackage{subeqnar}  
\usepackage{multicol}  
\usepackage{physprbb}  
\makeindex             

{\par\begingroup\addtolength{\leftskip}{#1}}%
{\par\endgroup} 
\begin{document}
\title*{Unified Model Photospheres\protect\newline 
for Ultracool Dwarfs of the Types  L and T
\thanks{
Proceedings of ``Ultracool Dwarfs: Surveys, Properties and Spectral
Classification''(August 12, 2000; Manchester) ed. H. R. A. Jones \&
I. Steele, Lecture Notes in Physics, Springer-Verlag}
}
\toctitle{Unified Model Photospheres for Ultracool Dwarfs 
of the \protect\newline Types  L and T}
\titlerunning{Unified Model Photospheres for Ultracool Dwarfs}
\author{Takashi Tsuji
}
\authorrunning{Takashi Tsuji}
\institute{Institute of Astronomy, School of Science, The University of Tokyo\\
2-21-1 Osawa, Mitaka, Tokyo, 181-0015 Japan}

\maketitle              

\def\la{\mathrel{\hbox{\rlap{\hbox{\lower4pt\hbox{$\sim$}}}\hbox{$<$}}}}
\def\ga{\mathrel{\hbox{\rlap{\hbox{\lower4pt\hbox{$\sim$}}}\hbox{$>$}}}}

\begin{abstract}
The presence of the two distinct groups of ultracool dwarfs (UCDs), `L' and `T' types, is now well established: L dwarfs are thought to be dusty while T dwarfs show strong bands of methane (CH$_4$) but little evidence for dust in their spectra. So far, different model sequences, which we referred to as the dusty model (or case B) and dust-segregated model (or case C) have been considered for L and T dwarfs, respectively.
We now propose instead that these two groups of UCDs may be understood
 as a temperature effect in a unique sequence of the model photospheres
 in which dust always exists but only in the restricted region (where $ 1800 \la T \la 2000$\,K) referred to as an active dust zone 
 (or this can be referred to as a dust cloud).
This is a natural consequence of considering not only dust formation but also 
its segregation process in the photosphere. By this model sequence, the
dust-column density in the observable photosphere first increases for
cooler objects and the infrared colours become redder from late M to L
dwarfs. On the other hand, the dust-column density in the observable 
photosphere decreases in the objects cooler than the latest L dwarfs as
the active dust zone moves to the optically thick region deep in the 
photosphere and the infrared colours turn blueward towards the coolest 
T dwarfs. In this way, the observed colours and spectra of UCDs through
L and T types can be explained consistently by a single grid of unified 
model photospheres. More generally, an important conclusion is that the 
photospheric dust formation is effective only in warmer deep regions. 
This is contrary to the general belief that dust forms in cooler surface 
regions.

{\bf key words:} dust-column density, L-type prototype GD165B, 
L/T transition objects, photospheric dust formation, quantitative 
spectroscopy, spectral classification, T-type prototype Gliese 229B
 
\end{abstract}

\section{Introduction}

Progress in observations of ultracool dwarfs (UCDs) has been quite substantial even in the short time since the discoveries of the prototypes such as GD165B ~\cite{b88} and Gliese 229B~\cite{n95}.  A large number of UCDs similar to GD165B have been discovered by the Two-Micron All-Sky Survey (2MASS)~\cite{k97} and the DEep Near-Infrared Sky (DENIS) survey~\cite{d97}. A sample of these L-type dwarfs is already large enough that its sub-types from L0 to L8 have been defined by Kirkpatrick et al.~\cite{k99b}~\cite{k00}. On the other hand, objects similar to Gliese 229B were more difficult to find, but it was not long before a dozen of cool brown dwarfs, referred to as methane dwarfs or T-type dwarfs, was discovered by the 2MASS~\cite{b99}~\cite{b00a}~\cite{b00b} and the Sloan Digital Sky Survey (SDSS)~\cite{s99}~\cite{t00b}. Finally, possible transition objects between L and T dwarfs were found by Leggett et al.\cite{l00a} and they were classified as early T dwarfs. This discovery confirmed that the L- and T-types form a single spectral sequence and may not be representing any kind of bifurcation.

The new spectral types L and T are added to the spectral types of O, B, A, F, G, K, \& M (with branching into R-N and S) established nearly a century ago. While the spectral sequence from O to M types is well understood as a temperature sequence by ionisation and dissociation theory, the problem is why dust apparently disappears in the cooler T dwarfs even if the spectral types L and T can be understood as an extension of the M type to the lower temperatures.
This article shows that the types L and T are indeed understood as a temperature sequence and that this is because the photospheric dust formation is effective only in the warmer deep region whose location depends on the effective temperature. This paper briefly reviews intriguing investigations that were carried out
prior to the solution (Sect.\,2) and discusses the resulting unified dusty model in some detail (Sect.\,3). Then it is shown that that the observed infrared colours of UCDs may be explained consistently by our grid of unified models and, further, that the infrared two colour diagrams provide a useful constraint on the location of the dust zone in the photosphere (Sect.\,4). Our models
are also applied to a preliminary analysis of the spectra of UCDs (Sect.\,5). While recognizing some inherent difficulties in probing dusty photospheres, 
we discuss some of the consequences of our new models (Sect.\,6).

\section{Modelling the Photosphere of the Ultracool Dwarf Stars}

Stellar photospheres are too hot for dust to form in general, but UCDs
are exceptional in that the thermodynamical condition of condensation is
well met in their photospheres. We had to confront how to treat dust in
modelling stellar photospheres for the first time for which some trial
and error were necessary. At first we considered dust formation and its
segregation in different models (Sect.\,2.1). Our initial attempt to
combine them in a single model was applied  to the specific case of
Gliese 229B (Sect.\,2.2).  We now show that dust formation and its
segregation process should be treated more consistently in a single
photospheric model and we finally have the unified model photosphere of
UCDs in which dust always exists but at different locations in L and T
dwarfs. We hopefully  conclude our exploratory stage of investigating
dusty model photospheres within the framework of the classical theory of stellar photosphere (Sect.\,2.3). 

\subsection{Dusty and Dust-Segregated Models : 1996 - 1998}
It was recognized that the condition of condensation is well fulfilled in the
photospheres of cool dwarfs~\cite{t96a}, but it was unknown how dust forms in
the photospheric environment. A problem is when nucleation begins after the
super-saturation ratio $S = p/p_{\rm sat}  (p_{\rm sat}$ is the saturation
vapour pressure) exceeds unity. We assumed two extreme cases: One is a dust-free model in which dust is not formed even if $ S > 1$ and the other is a dusty model in which dust forms as soon as $S$ exceeds unity. It is found that the dusty models explain the observed characteristics of late M dwarfs~\cite{t96a}~\cite{j97} as well as L-type prototype GD 165B~\cite{t96b}. In this case, we assume that the small dust grains formed at relatively high temperatures remain well mixed with the gaseous components.  
On the other hand, the genuine brown dwarf Gliese 229B discovered by Nakajima et al.~\cite{n95} shows little evidence for dust but could be explained rather well by our dust-free model developed before the discovery of the brown dwarf~\cite{t95}.
We interpreted this result as due to the segregation of dust which once formed and grew too large to be sustained in the photosphere of the cool brown dwarf~\cite{t96b}.

The preceding results are easily consistent with the classical nucleation theory according to which the dust growth cannot start before its radius $ r_{\rm gr} $ reaches the critical radius $r_{\rm cr}$, where the Gibbs free energy of condensation shows the maximum. The dust grains with  $ r_{\rm gr} < r_{\rm cr}$ are thermodynamically unstable, in the sense that the dust grains formed will soon dissolve and {\it vice versa}.
In other words, such small dust grains are in detailed balance with the gaseous mixture and hence can easily be sustained in the photosphere. 
For this reason, such small grains that failed to be the stable large grains, play an important role as sources of opacity in the photosphere.
On the other hand, the dust grains with $ r_{\rm gr} > r_{\rm cr}$ are stable and grow larger and larger.  Such large grains, however, will segregate from the gaseous mixture and no longer be sustained in the photosphere. Thus, large grains may be formed, but they may precipitate below the photosphere and will play little role as sources of opacity.   
Then, we distinguished three cases of $ r_{\rm gr} =0 $,
$ r_{\rm gr} \la r_{\rm cr}$ and $ r_{\rm gr} > r_{\rm cr}$ which we referred to as case A (super-saturated case), B (dusty case) and 
C (dust-segregated case), respectively~\cite{t00a}.
We hoped that L and  T dwarfs could be accounted for
by the dusty models (case B) and dust-gas segregated models (case C), respectively. One difficulty, however, was the large excess of the optical flux predicted by the dust-segregated model compared with the observation of Gliese 229B~\cite{g98}.

\subsection{A Hybrid Model: 1999}

An initial motivation to consider a hybrid model, which consists of the warm dust in the deeper layer and cool volatile molecules in the upper layer, was to explain a large flux depression in the optical spectrum of Gliese 229B~\cite{t99b}. In fact, if dust plays some roles in depressing the optical flux, only the dust deep in the photosphere will work for this purpose, since the dust in the surface region will mask other prominent spectral features such as due to CH$_{4}$ and H$_{2}$O. At the same time, however, it was noticed that the effect of the pressure-broadened wings of alkali metals depresses the optical flux significantly~\cite{t99b}~\cite{b00c}. 
Then, the dust is not necessarily called for to explain the optical flux depression.

Nevertheless, we will show that the warm dust deep in the photosphere will have noticeable observable effects in L and early T dwarfs (Sect.\,2.3). 
In fact, the significance of the hybrid model is not necessarily to explain the optical flux depression, but rather it involves an important idea to be developed to a unified model of UCDs.
Our previous dusty (case B) or dust-segregated (case C) models were based on the assumptions that the dust once formed remains throughout the photosphere in L dwarfs or segregation process takes place throughout the photosphere in T dwarfs, respectively. But, it is difficult to understand why the different cases are realized for the same physical condition that may be found somewhere in the photospheres of L and T dwarfs. Although these cases B and C models could explain some characteristics of a few selected  M, L and T dwarfs, new observations on a larger sample revealed some inconsistencies and  suggested a  more realistic model somewhere in between these two extreme cases~\cite{t99a}. 
Our hybrid model may already be suggesting a way to relax these issues.  
  
\subsection{The Unified Model : 2000}
 Once again, we remember how dust forms in the photospheric condition. Dust forms as soon as temperature is lower than the condensation temperature ($T_{\rm cond} $) but the dust will soon grow too large at slightly lower temperature, say  $T_{\rm cr}$ (critical temperature), when the dust size reaches its critical radius $ r_{\rm cr} $.  Thus, in the region with  $T \la T_{\rm cr}$ in the photosphere, dust will be large enough to segregate from the gaseous mixture  and soon precipitate below the photosphere. Only in the region with $ T_{\rm cr} \la T \la T_{\rm cond} $, the dust grains will be small enough ( $ r_{\rm gr} \la r_{\rm cr} $) to be sustained in the photosphere and it is such small dust grains that play 
an important role as opacity sources.  We refer to this region with 
$ T_{\rm cr} \la T \la T_{\rm cond} $  as an active dust zone. 
Thus, the dust effectively exists only in the relatively warm region deep in 
the photosphere, 
 and this means that a dust layer or a cloud is formed in 
the photosphere.
This is a natural consequence of consistently considering not only dust
formation but also its segregation process. 
In this way, the model with the warm dust 
 layer deep 
in the photosphere can be constructed anew based on a clear physical basis rather than as a ``hybrid'' of  any preceding models as its constituents.
For this reason, it is not appropriate to refer to such a model with the active dust zone as a hybrid model and may be referred to as a unified model to distinguish it from the previous models.

Now, a major problem is to find the temperatures that define the active
dust zone. The condensation temperature $ T_{\rm cond} $ is easily found
from the thermochemical computation during the model construction to be
$ T_{\rm cond} \approx 2000$\,K for  corundum and iron, which first form
in the photospheres of UCDs~\cite{t96a}. On the other hand, the critical
temperature $ T_{\rm cr} $ is more difficult to find. This should in
principle be determined from the detailed analysis of the dust-gas
segregation process, but it still seems to be premature to solve this
problem theoretically. Instead, we treat $ T_{\rm cr} $ as a free
parameter to be found empirically.  Actually, we find that $ T_{\rm cr}
$ can be constrained well by the infrared colours which showed red
limits at late L dwarfs~\cite{k00} and we find it to be $ T_{\rm
cr}\approx 1800$\,K (Sect.\,4). 
 The critical radius  $r_{\rm cr} $ itself is more
difficult to estimate, but astronomical grains of
 0.01 $\mu$m or smaller are known and we may 
assume that $r_{\rm cr} < 0.01 \,\mu$m. On the other hand, 
 mass absorption coefficients of dust grains
depend little on the grain sizes so long as the grains are smaller
than about 0.01 $\mu$m, and the value of  $r_{\rm cr} $
gives little direct effect on our actual modelling.

 It is to be noted that the active dust zone can be found in all the cool dwarfs with $ T_{\rm eff} \la 3000$\,K. For objects with very low $ T_{\rm eff} $ near 1000\,K, this active dust zone is situated too deep in the photosphere (where $\tau_{\rm Ross} > 1$ , since $ T \approx T_{\rm eff}$ at $\tau_{\rm Ross} \approx  1$) and this should be the reason why dust apparently shows little observable effects in cool T dwarfs. On the other hand, for the relatively warm objects with $ T_{\rm eff} $ above about 1500\,K (see Sect.\,4), the active dust zone is situated nearer to  the surface ($\tau_{\rm Ross} < 1$) and this explain why L dwarfs, whose $ T_{\rm eff} $ may be higher than about 1500\,K~\cite{r99}, appears to be dimmed by dust.
The case of the early T dwarfs recently discovered~\cite{l00a}, the active dust zone may just be situated near the optical depth unity and thus these objects with $ 1200 \la T_{\rm eff}\la 1400$\,K may represent the L/T transition objects.

\begin{figure}[b]
\begin{center}
\includegraphics[width=.7\textwidth]{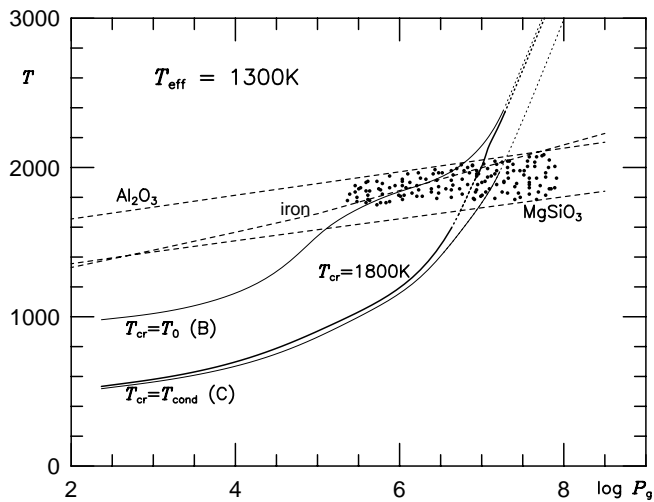}
\end{center}
\caption{Model photospheres of $T_{\rm eff} =  1300$\,K  
(solar metallicity, log $g$ = 5.0 \& $v_{\rm micro} = 1$\,km\,s$^{-1}$). 
The heavy line represents the unified model with $T_{\rm cr} =1800$\,K  while the thin lines the limiting cases of  $ T_{\rm cr} = T_{0}$ (case B) and 
$ T_{\rm cr} = T_{\rm cond}$ (case C). The solid and dotted lines represent radiative and convective zones, respectively. The condensation lines of corundum (Al$_{2}$O$_{3}$), iron and  silicate (MgSiO$_{3}$) are shown by the dashed lines.  The active dust zone is indicated by the dotted area.}
\label{eps1}

\end{figure}

\section{Physical Structures of the Unified Models}

We will discuss some details of our unified model for the case of $ T_{\rm eff} = 1300$\,K  as an example  in Fig.\,1. The active dust zone is shown by the dotted area, the upper boundary of which is defined by the condensation line of corundum (Al$_2$O$_3$) shown by the dashed line. Before the active dust zone terminates at $T \approx T_{\rm cr} \approx 1800$\,K, iron (Fe) condenses at its condensation line. However, enstatite (MgSiO$_3$) may form outside the active dust zone and this means that enstatite will segregate as soon as it is formed. This may be possible, since enstatite will easily form with corundum 
and/or iron as the seed nuclei and grow rapidly. By the same reason, other solid species that may form at lower temperatures will precipitate as soon as they are formed. For this reason, only the dust species formed at relatively high temperatures above about
$T_{\rm cr} \approx 1800$\,K work as the active dust ({\it i.e.} as source
of opacity) and hence give significant effect on the photospheric structure.
This fact may simplify the construction of models since it is enough to consider only the high temperature condensates such as corundum and iron as sources of opacity. In the active dust zone, the temperature gradient is quite steep  because of the high opacity of the dust and the model is convectively unstable near $ T_{\rm cr}$. For this reason, our unified model shows the outer and inner convective zones separated by an intermediate or detached radiative zone (Fig.\,1), as discussed in the case of the hybrid model for Gliese 
229B~\cite{t99b}.

In Fig.\,1, the resulting structure of our new model is found between the previous dusty (cases B) and dust-segregated (case C) models as can be expected. In fact, we assumed that the dust grains once formed remain small enough throughout the photosphere and never precipitate up to the stellar surface in our previous dusty model (case B). 
This is equivalent to have assumed $ T_{\rm cr} = T_{0}$ 
($T_{0}$: surface temperature). On the other hand, we assumed that the dust grains will precipitate as soon as they are formed in our previous 
dust-segregated model (case C) and this is equivalent to have assumed 
$ T_{\rm cr} = T_{\rm cond}$. Thus, our previous models of cases B and C represent the extreme limiting cases of our new model. 
Then, the active dust in our new model works to heat up the photosphere near the active dust zone but not so much as in our previous case B model in the upper layer. On the other hand, volatile molecules work as dominant sources of opacity after dust has precipitated in the layer above $ T = T_{\rm cr}$ in our new model and the photosphere is cooled appreciably by the cooling effect of the volatile molecules as in our case C model. It is to be noted, however, that the photospheric structure of the unified model approaches to that of the previous case B in the region below the active dust zone.

\begin{figure} [b]
\begin{center}
\includegraphics[width=0.95\textwidth]{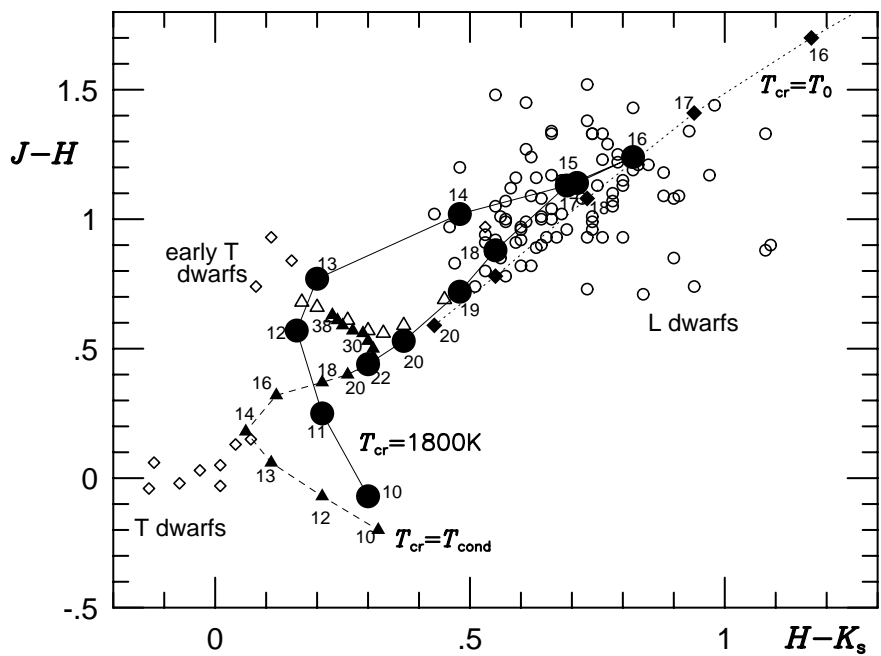}
\end{center}
\caption{ ($J - H$, $H - K_{s}$) diagram. Observed colours of M, L and T dwarfs are shown by the open triangles~\cite{l92}, 
circles~\cite{k99b}~\cite{k00} and 
squares~\cite{m96}~\cite{s99}~\cite{b99}~\cite{b00a}~\cite{b00b}~\cite{t00b} 
~\cite{l00a}, respectively. Predicted colours based on our unified models with $T_{\rm cr} \approx 1800$\,K are shown by the filled circles while those of the dusty (case B) and dust-segregated (case C) models by the filled squares and triangles, respectively. The numbers attached are $T_{\rm eff}$'s in unit of hundred Kelvin (the steps of $T_{\rm eff}$'s are 100 or 200\,K)
}
\label{eps2}
\end{figure}

\section{Colours}

The infrared colours of UCDs are not necessarily redder for cooler objects but turn to blue in T dwarfs after passing the red limits around late L dwarfs~\cite{k00}. These observations indicate that there should be an additional parameter other than $T_{\rm eff}$ in determining the colours.
This parameter should be related to the dust in the photosphere and we identify it with $T_{\rm cr}$. We will show that the rather complicated behaviours (Figs.\,2 \& 3) of the infrared colours are well understood by our unified models characterized by  $T_{\rm eff}$ as well as by $T_{\rm cr}$ and that the effect of these two parameters can be separated on the infrared two-colour diagrams.

\subsection{($J - H$, $H - K_{s}$) Diagram}

The $J-H$ and $ H-K_{s}$ are best observed for a large sample of UCDs.
We reproduced the observed colours of M, L and T dwarfs including the early T-type, by the open triangles, circles and squares, respectively, in Fig.\,2. With the discovery of the early T dwarfs~\cite{l00a}, which may be the L/T transition objects, it now appears that the UCDs show a continuous loop counter-clockwise on the ($J - H$, $H - K_{s}$) diagram from M to T dwarfs via L dwarfs. Also, an interesting feature is the presence of the red limits for the infrared colours~\cite{k00}. 

In Fig.\,2, the predicted colours based on our previous case B models (filled squares) could explain  the very red colours of some L dwarfs, but could not the presence of the observed red limits of $J - H$ and $H - K_{s}$. On the other hand, our case C models (filled triangles) could explain the very blue colours of T dwarfs (but predicted $H - K_{s}$'s are too red by about 0.3 mag. as compared with observed and this is due to the difficulty to predict the $K$ flux accurately by our models as will be noted in Sect.\,5.2), but could not  the reddening of $J - H$ towards early T dwarfs. On the other hand, our new model (filled circles) roughly explains the general trend of the observed colours through M, L, L/T and T dwarfs by a single grid of the model photospheres with $T_{\rm cr} \approx 1800$\,K.

\begin{figure} [b]
\begin{center}
\includegraphics[width=0.95\textwidth]{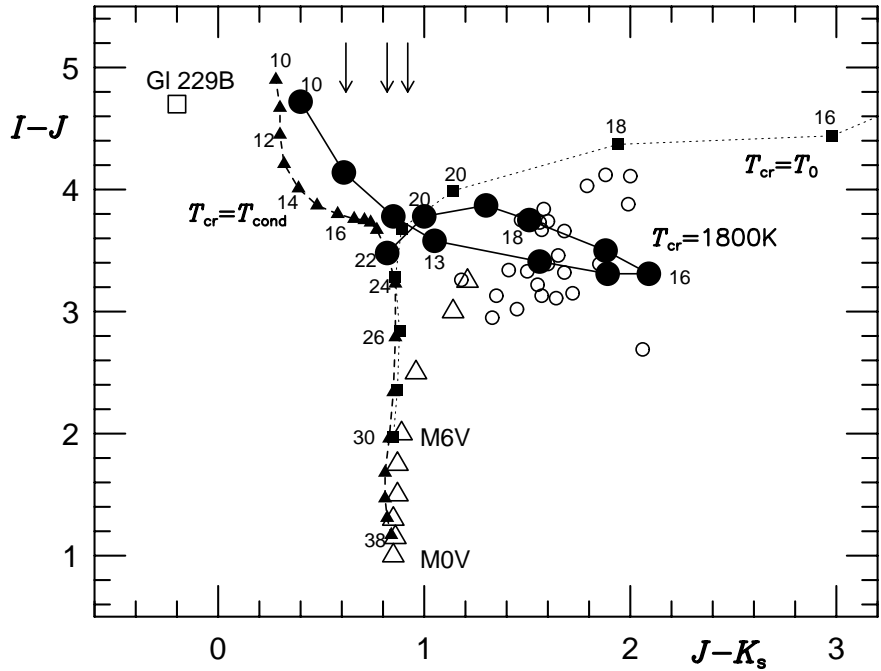}
\end{center}
\caption{ 
($I - J$, $J - K_{s}$) diagram. The arrows indicate the
values of $J - K_{s}$ for three L/T transition objects~\cite{l00a}.
See Fig.\,2 legend for details
}
\label{eps3}
\end{figure}

The redness of the infrared colours is essentially determined by the mass-column density  of the active dust in the observable photosphere.  Since the active dust zone is within the optically thin
 regime in the relatively warm objects, the dust-column density in the observable photosphere first increases towards cooler objects from late M to L dwarfs and the infrared colours show reddening in agreement with observations. However, the dust-column density in the observable photosphere decreases towards the coolest objects even if the mass-column density of the active dust zone itself increases, because it now penetrates into the optically thick regime in the cooler objects. Our model grid predicts that this takes place at  $ T_{\rm eff} \approx 1600$\,K (Fig.\,2), which may correspond to the latest L dwarf~\cite{r99}.  This explains the presence of the red limits in the ($J - H$, $H - K_{s}$) colours and the bluer colours of L/T transition objects as well as of T dwarfs.

The observed red limits of ($J - H$, $H - K_{s}) \approx (1.3,0.8)$~\cite{k00}  are well explained by our model grid based on $ T_{\rm cr} = 1800$\,K which predicts the red limits of ($J - H$, $H - K_{s}) \approx (1.2, 0.8)$ at $ T_{\rm eff} \approx 1600$\,K (Fig.\,2). The mass-column density of the active dust zone in the observable photosphere should also be larger for the lower $ T_{\rm cr}$, since this means that the active dust zone extends outward and the red limits will still be redder. In fact, our grid based on $ T_{\rm cr} = 1600$\,K  predicts the red limits of ($J - H$, $H - K_{s}) \approx
(1.5, 1.1)$.  On the other hand, another grid based on the higher
$T_{\rm cr} = 1900$\,K predicts  the red limits of  
($J - H$, $H - K_{s}) \approx (1.0, 0.5)$.
Thus, the value of $ T_{\rm cr} = 1800$\,K can be regarded as being well constrained.

\subsection{($I - J$, $J - K_{s}$) Diagram}
We  show the observed and predicted $I - J$ and $J - K_{s}$ colours in Fig.\,3. Here, the observations show bifurcation to the red (L dwarfs: open circles) and blue (Gliese 229B: open squares) sequences. The bifurcation could apparently be explained by our previous case B (filled squares) and C (filled triangles) models. However, our unified models (filled circles) indicate a possibility that these sequences are in fact understood as a single sequence. The observed  $J - K_{s}$ colours of the three L/T transition objects~\cite{l00a} shown by the arrows in Fig.\,3 confirm that the observed colours form a single continuous loop in the ($I - J$, $J - K_{s}$) diagram.
The reason for this is essentially the same as for ($J - H$, $H - K_{s}$) figure. Also, the observed red limit of $J - K_{s} \approx 2.1$ is well reproduced by our model grid based on $T_{\rm cr} = 1800$\,K (Fig.\,3), while it could not be predicted at all by our previous cases B and C models.
For comparison, the predicted red limits are $J - K_{s} \approx 2.6$ and 1.5  for our grids based on $T_{\rm cr} = 1600$ and 1900\,K, respectively. Again, our choice of $T_{\rm cr} = 1800$\,K is well justified by the ($I - J$, $J - K_{s}$) diagram. However, our predicted $J - K_{s}$'s are too blue for the late M and early L dwarfs while too red for the coolest T dwarf Gliese 229B (also see 
Fig.\,5). Probably, something may be still missing in our opacity data.

\begin{figure} [b]
\begin{center}
\includegraphics[width=1.0\textwidth]{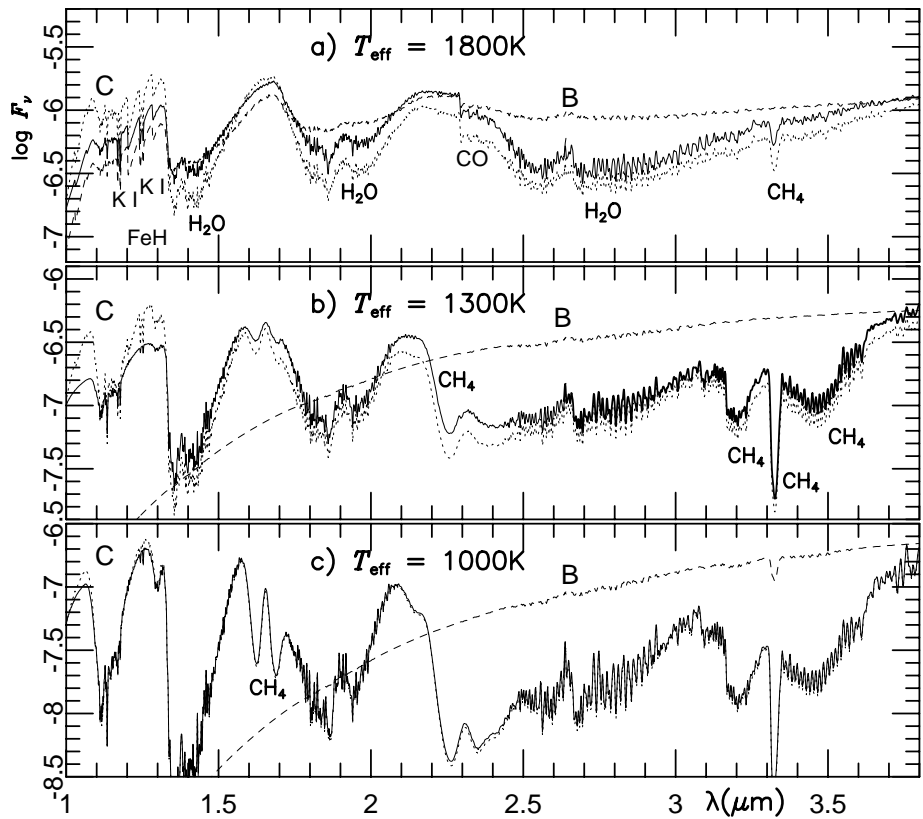}
\end{center}
\caption{ 
The predicted spectra (in erg/cm$^2$/sec/Hz) based on our models 
($T_{\rm cr} = 1800$\,K) are shown by the solid lines while those of the limiting cases of $ T_{\rm cr} = T_{0 }$ (dusty case B) and $ T_{\rm cr} = T_{\rm cond}$ (dust-segregated  case C) by the dashed and dotted lines, respectively.
(\textbf{a}) $ T_{\rm eff} =1800$\,K. 
(\textbf{b}) $ T_{\rm eff} =1300$\,K. 
(\textbf{c}) $ T_{\rm eff} =1000$\,K 
}
\label{eps4}
\end{figure}

\section{Spectra}

In our unified models, only the dust in the active dust zone located within the observable photosphere gives some observable effects, but even such a small amount of dust gives appreciable effects on spectra as well as on colours because of the very large extinction of dust. We first show a general characteristics of the infrared spectra based on our new grid of the unified models (Sect.\, 4.1). Then, we rediscuss the prototype of T-type Gliese 229B (Sect.\, 4.2) and that of L-type GD165B (Sect. 4.3) by our models. 

\subsection{Predicted Spectra of the Unified Models}

The predicted spectra (in $F_{\nu}$ unit) based on our new models shown in Fig.\,4 can  be seen to be well consistent with the infrared colours discussed in Sect.4.    
For example, the mass-column density of the active dust zone in the observable photosphere is increasing in our model of $ T_{\rm eff} = 1800$\,K 
(Fig.\,4a) and the effect of dust on the spectrum is appreciable.
The effect of dust extinction on the $J$ band is largest in case B, shows no effect in case C and it is just intermediate between these extreme cases in our new model, as is the $J-H$ (Fig.\,2). At the same time, the dust also contributes to heat the photosphere and the $K$ band region shows the opposite tendency because the effect of the H$_2$ collision-induced absorption (CIA) which is less important in the warmer photosphere of the larger dust-column density. In the case of $ T_{\rm eff} = 1300$\,K (Fig.\,4b), the mass-column density of the active dust zone itself still increases, but the part in the observable photosphere decreases since part of it is now below the optical depth unity. For this reason, the effect of dust is only modest, but the $J$ flux still suffers considerable extinction by the dust resulting in the reddening of $J-H$ (Fig.\,2).
Finally, in the model of $ T_{\rm eff} = 1000$\,K (Fig.\,4c), the predicted  infrared spectrum based on our new model differs little from that of the case C, as are the infrared colours (Figs.\,2 \& 3). This is because the active dust zone is situated below the observable  photosphere of this very cool model.   

It is interesting to see in our unified model that the  $Q$ branch of the CH$_{4}$ $\nu_{3}$ fundamentals appears by $ T_{\rm eff} = 
1800$\,K and that the weaker combination bands at 1.6 and 2.2\,$\mu$m are strong by $ T_{\rm eff} = 1300$\,K. 
These results find observational support in the recent detections of
CH$_{4}$ bands in L dwarfs reported in this meeting by Geballe and by Noll as well as in the early T dwarfs by Leggett et al.~\cite{l00a}. The 
CH$_{4}$ $\nu_{3}$ fundamentals appeared at $ T_{\rm eff} = 1800$\,K in
our previous model C (Fig.\,4 in~\cite{t00a} in which CH$_{4}$ bands
were probably overestimated by the use of the smeared out CH$_{4}$
opacity), but we did not think this to be serious since we thought that our
previous model C cannot be applied to L dwarfs. This conclusion remains
unchanged even if  CH$_{4}$ bands can be predicted  by this
model. However, our previous dusty model (case B) never predicts the
CH$_{4}$ bands (Fig.\,4a) and the recent detection of CH$_{4}$ in L dwarfs
completely ruled out the possibility of the simple dusty models for L
dwarfs. Thus, our new model provides a distinct possibility to explain 
the presence of methane as well as of dust in L dwarfs consistently.

In the present work, we use the line databases HITEMP~\cite{r97} for  H$_{2}$O and GEISA~\cite{j99} for CH$_{4}$ $\nu_{3}$ fundamentals, but we still use the smeared-out band models for CH$_{4}$ combination bands as well as for FeH. The present 
CH$_{4}$ linelist, however, is valid only at low-temperatures and its extension to the higher temperatures is urgently needed.

\subsection{ The Spectrum of the T Dwarf Prototype Gliese 229B}

We compare the observed  spectrum of Gliese 229B  by Geballe et 
al.~\cite{g96} (calibrated by Leggett et al.~\cite{l99}) and by Oppenheimer et al.~\cite{o98}, with the predicted ones based on our new model of $T_{\rm eff} = 1000$\,K in Fig.\,5. To show the effect of 
$T_{\rm cr}$, we show two models: One with $T_{\rm cr} = 1800$\,K which we now believe to be the best  (see Sect.4) and the other with $T_{\rm cr} =
 1600$\,K which is close to the value we applied to Gliese 229B in our previous analysis~\cite{t99b}.
Inspection of Fig.\,5 reveals that the case of $T_{\rm cr} = 1800$\,K gives an overall better fit than the case of $T_{\rm cr} = 1600$\,K.
This fact confirms that our choice of $T_{\rm cr} = 1800$\,K is acceptable for the coolest T dwarfs as well. But we identify two major discrepant regions. First, the predicted flux appears to be higher than the observed at the $K$ band region and this is the reason why the predicted  $H - K_{s}$ and  $J - K_{s}$ are too red (Sect.\,4).
 Here, our opacity data (e.g. CH$_{4}$) may not be perfect and some unknown sources may also be possible at the very low temperatures.
Second, the optical flux cannot yet be explained quantitatively and one possibility may be to improve the broadening theory of strong alkali metal lines as suggested by Burrows et al.~\cite{b00c}.

\begin{figure}[t]
\begin{center}
\includegraphics[width=1.0\textwidth]{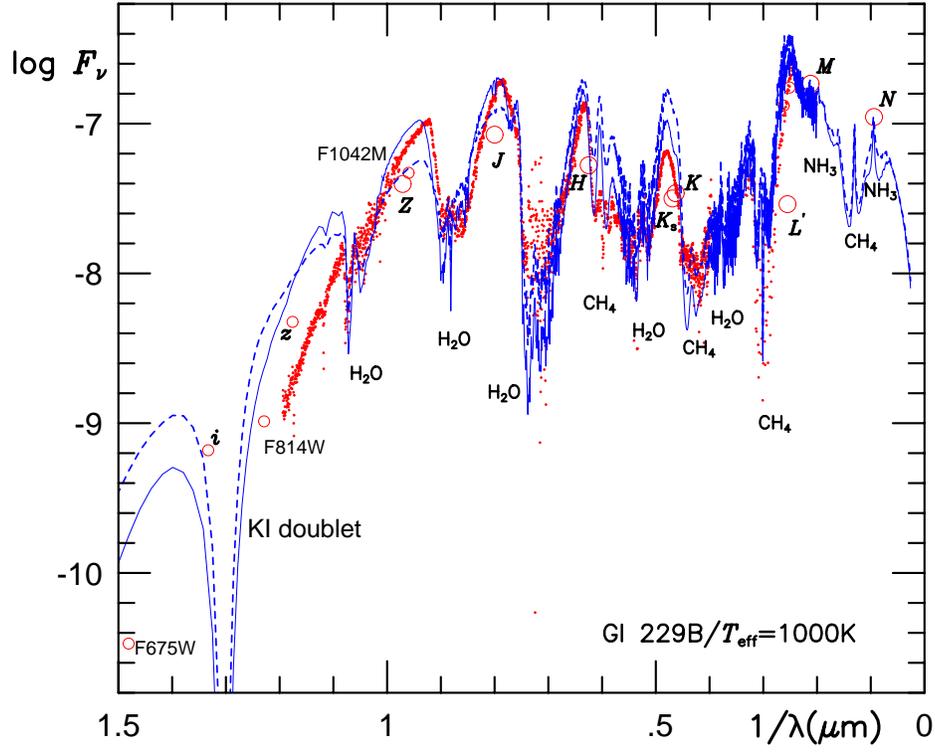}
\end{center}
\caption{ Observed spectrum~\cite{g96}~\cite{l99}~\cite{o98} and
 photometry data~\cite{m96}~\cite{g98}
of Gliese 229B are shown by the dots and open circles, respectively. 
The predicted spectra (in erg/cm$^2$/sec/Hz) based on our unified models 
of $ T_{\rm eff} = 1000$\,K are shown by the solid and dashed lines
for the cases with $T_{\rm cr} = 1800$\,K and 1600\,K, respectively 
}
\label{eps5}
\end{figure}

\begin{figure}[t]
\begin{center}
\includegraphics[width=1.0\textwidth]{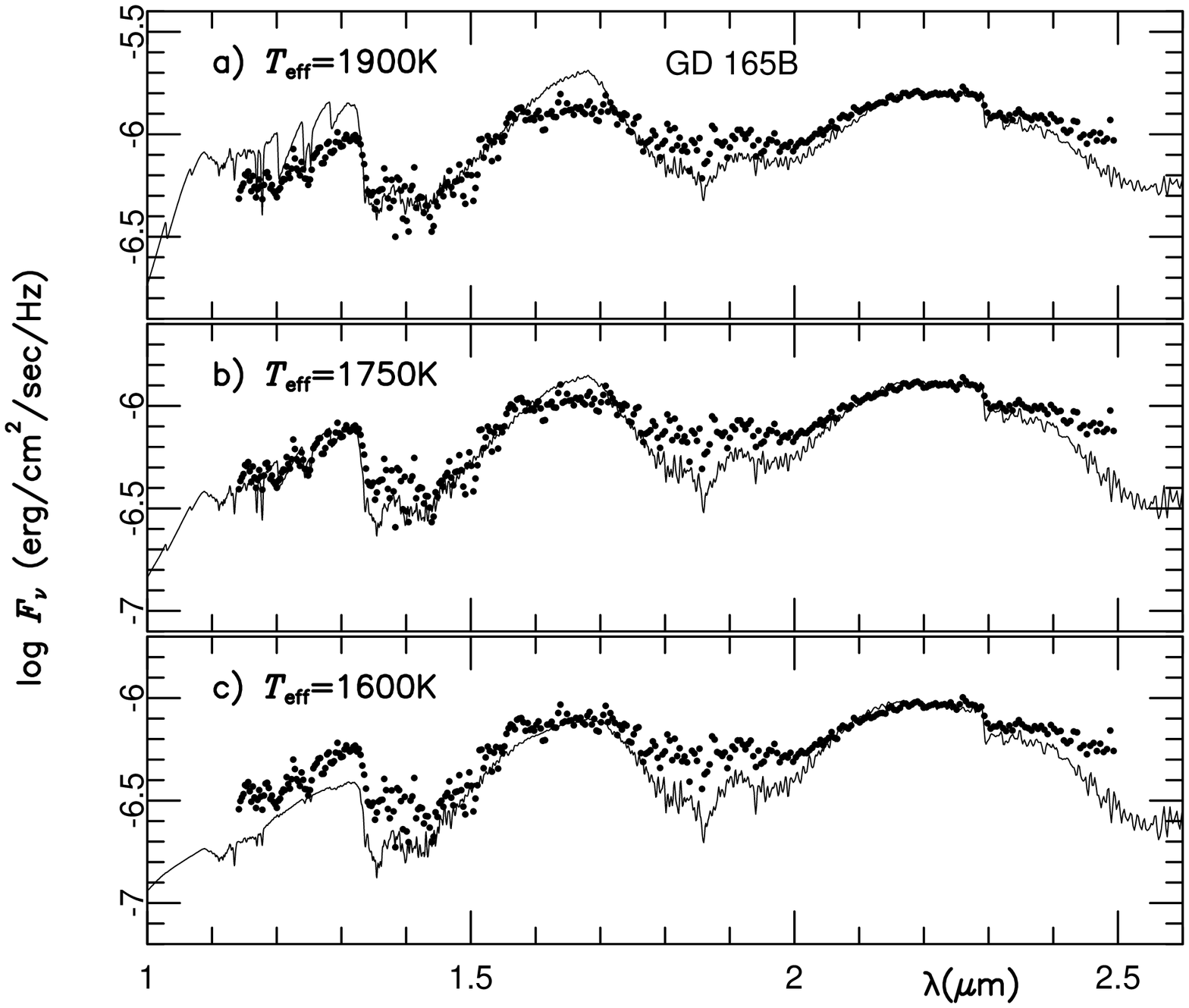}
\end{center}
\caption{ Observed spectrum of GD 165B~\cite{j94} shown by the dots is
fitted (first at the $K$ band region) to the predicted ones  based on our 
unified models ($T_{\rm cr} = 1800$\,K) shown by the solid lines. 
(\textbf{a}) $ T_{\rm eff} =1900$\,K. 
(\textbf{b}) $ T_{\rm eff} =1750$\,K. 
(\textbf{c}) $ T_{\rm eff} =1600$\,K 
}
\label{eps6}
\end{figure}

Our previous choice of $T_{\rm cr} = 1550$\,K~\cite{t99b} was largely biased towards  explaining  the observed spectrum of Gliese 229B in the optical region, 
for which  the  predicted flux based on $T_{\rm cr} = 1600$\,K in fact shows
a better fit (e.g. $i$-flux in Fig.\,5). However, we now recognize that 
$ T_{\rm cr} $ cannot be determined from such a very cool object alone whose active dust zone gives little observable effect.
In fact, our new model  with  $T_{\rm cr} = 1800$\,K predicts almost the same spectrum as our previous dust-segregated model for $T_{\rm eff} = 
1000$\,K (Fig.\,4c).    For this reason, it may not be possible to prove nor to disprove the presence of the warm dust deep in the photosphere by the analysis     
such as done on the spectrum of SDSS 1624~\cite{l00b}~\cite{n00}.
Thus, the presence of the warm dust proposed for Gliese 229B~\cite{t99b} cannot be confirmed by this object itself, but can be deemed as well established by our analysis of a larger sample of UCDs (e.g. Sect.4).

\subsection{The Spectrum of the L Dwarf Prototype GD 165B}

Our old dusty model (case B) already explained the observed spectrum of GD165B rather well while our dust-segregated model (case C) could not~\cite{t00a} and GD165B may in fact be remembered as the first object in which the presence of dust in the photosphere was recognized~\cite{t96b}. 
But this result should be somewhat fortuitous in view of our new models and  we rediscuss the observed spectrum of
GD165B by Jones et al.~\cite{j94} in Fig.\,6.
It is clear that our new models explain the observed spectrum as well, especially if we assume $ T_{\rm eff} = 1750$\,K. 
It is to be remembered that the best fit with our previous dusty models was obtained for $ T_{\rm eff} = 1800$\,K~\cite{t96b} and another independent analysis  suggested $ T_{\rm eff} = 1900$\,K~\cite{k99a}. Thus, the effect of the new model is to lower the estimated $ T_{\rm eff}$. 
Although all these models applied so far to GD165B provide more or less similar good fits, our new model is physically more reasonable and, moreover, our analyses of the infrared colours for
a large sample of UCDs  provides definite evidence for our new models 
(Sect.\,4).
Thus, GD165B may have $ T_{\rm eff} \approx 1750$\,K and now be closer to the
substellar regime although its substellar nature may still
depend on the details of the evolutionary models~\cite{b97}~\cite{c00}. 
This result, however, may not be completely
free from the difficulty to be discussed in Sect.\,6.3, even though 
$ T_{\rm cr}$ is relatively well estimated for the L dwarf regime. 

\section{Discussion}

An important conclusion on dust formation in the photospheric environment is that it is effective only in the warmer deep region and not in the cooler surface region (Sect.\,6.1).  Once this simple principle is realized, we can construct reasonably realistic model photospheres for UCDs and the new spectral types L and T can consistently be interpreted as a temperature sequence (Sect.\,6.2). However, the details of spectra depend on the dust-column density in the observable photosphere and fully quantitative analyses of the spectra of dusty photospheres should have some inherent difficulties.
Furthermore, it is difficult to prove the presence of dust if it is below the observable photosphere as spectroscopic diagnosis is impossible for anything below the photosphere (Sect.\,6.3).

\subsection{Dust Formation in the Photospheric Environment}

Unlike the case of cool giant stars where dust forms in the outflow, dust in cool dwarf stars forms in the static photosphere and there should be fundamental differences in the dust formation mechanisms in low and high luminosity stars. One problem is how dust could be sustained in the photosphere of dwarf stars for a long time.
After considering not only dust formation but also its segregation process, we arrive at a conclusion that dust formed can be sustained only in the region near the condensation temperatures in the photosphere (Sect.\,2.3). This means that dust effectively exists in the rather warm region relatively deep in the photosphere contrary to the general belief that dust is more abundant in the cooler surface region. This conclusion may apply not only to UCDs but also to the dust formation
in the photospheric environment in general, including extrasolar giant 
planets (hot Jupiters), proto stars,  accretion disks etc.

This conclusion on photospheric dust formation is confirmed by the fact 
that our new models based on this assumption explain the observed 
colours (Sect.4) and spectra (Sect.5) rather well. On the other hand, 
our previous dusty (case B) and dust-segregated (case C) models,
which represent the extreme limiting cases of $ T_{\rm cr} = T_{0 } $ 
($T_{0 } $: surface temperature) and $ T_{\rm cr} = T_{\rm cond } $, 
 respectively,  could not explain the observed colours of UCDs consistently
(Figs\,2 \& 3). 
 Thus, it is clear that our previous models are not realistic enough to be 
used for interpreting observed data and should no longer be used.
The model photospheres of UCDs by other authors are  
also subject to the same criticism. For example, the DUSTY and COND
models discussed recently by Chabrier  et al.~\cite{c00} correspond to the 
limiting cases of $ T_{\rm cr} = T_{0 }$  and $ T_{\rm cr} = T_{\rm cond}$,
respectively. The photospheric  models of cool brown dwarfs
including  giant planets by Burrows et al.~\cite{b97}  also essentially 
assumed $ T_{\rm cr} = T_{\rm cond}$ throughout. 
In the case of the coolest brown dwarfs, where dust exists deep in the photosphere, the models with $ T_{\rm cr} = T_{\rm cond}$ give essentially the same emergent spectra as the unified models. However, dust deep in the photosphere gives considerable effect on the structure of the inner photospheres and hence on the boundary condition for the interior models. 
 
\subsection{ Stellar Spectral Classification Extended to UCDs}

The spectral classification of L dwarfs by Kirkpatrick 
et al.~\cite{k99b}~\cite{k00} is based on a large sample of the far-red 
spectra of UCDs (0.63 -- 1.01 $\mu$m).
 The resulting spectral subclasses show good 
correlations with the infrared colours, although the correlations are not 
monotonic especially if T dwarfs are considered, but show a red 
limit at the latest L dwarfs~\cite{k99b}~\cite{k00}. 
We have shown that the infrared colours of UCDs are essentially controlled by 
the mass-column density of the dust in the observable photosphere which first 
increases towards lower $T_{\rm eff}$ but shows a maximum at about $T_{\rm eff}
 \approx 1600$\,K and then decreases towards the coolest T dwarfs (Sect.\,4). 
Thus, the infrared colours are well correlated with stellar temperatures 
and this fact confirms that the spectral classification by Kirkpatrick 
et al.~\cite{k99b}~\cite{k00} also represents the temperature sequence.
It is remarkable that the spectral classification by Kirkpatrick 
et al.~\cite{k99b}~\cite{k00} done on a purely empirical basis
reflects the photospheric structure of UCDs so well.
We also propose that L and T types represent the objects in which
the dust-column density in the observable photosphere is increasing and
decreasing respectively. Since dust is the major ingredient in determining
the photospheric structure, this interpretation of L and T types should
be more fundamental than the use of methane which is now observed both in L and T
dwarfs (Sect.\,5.1). Anyhow, it may be reasonable to have divided 
UCDs into L and T types.

So far, we have not as yet analysed directly the spectral features 
used as classification criteria~\cite{k99b}~\cite{k00}, 
including oxides such as TiO and VO, 
hydrides such as FeH and CrH and neutral alkali metals.
The absorption  bands due to the refractory compounds should be formed
below the active dust zone where refractory elements are not yet depleted in dust. This dust free zone in the observable photosphere
({\it i.e.} in $\tau < 1$) shrinks as the active dust zone moves towards
deeper region in cooler L dwarfs and this is one reason why
TiO and VO are weaker in cooler L dwarfs. Further, the mass-column 
density of the active dust above this molecule-dominated region
is larger for cooler L dwarfs and molecular bands suffer larger
extinction by the dust. Besides these two major effects
of dust, the observed band strengths also depend on
the gas phase chemical equilibrium. For example,  TiO and VO
attain their maximum abundances already in late M dwarfs while 
hydrides such as FeH and CrH may still be increasing in the L dwarf
regime because of their lower dissociation energies. 
For this reason, hydrides are well observed in L dwarfs. On the
contrary, alkali metals may be abundant in the region above the
active dust zone and may be stronger for cooler objects including
T dwarfs. Thus, atomic and molecular spectra will provide
abundant information on the structure of the dusty photosphere.

The spectral sequence of O -- M can be understood as a temperature
sequence by considering ionisation and dissociation in gaseous mixture.
The new spectral types L and T  can be understood as a temperature sequence 
in which the dust forming region moves from the optically-thin region in 
L dwarfs to the optically-thick region in T dwarfs. 
Thus, the stellar spectral classification including the  L- and T-types 
can be understand as a single 
sequence of temperature and  
a large variety of spectra from OB stars to brown dwarfs including
L and T types can be interpreted 
by a simple thermodynamics including dust condensation 
and segregation in addition to ionisation and dissociation.

\subsection{Spectroscopic Diagnosis of Dusty Photospheres}
The observed colours and spectra of UCDs can 
in principle be interpreted by our unified models, but
this does not imply that the quantitative spectroscopy
of  UCDs with the same accuracy as in non-dusty stars can be possible.
The inherent difficulty is that the spectra of dusty objects depend strongly on the dust-column density in the observable photosphere (Sect.\,5). Since dust shows no observable feature by itself, 
it is very difficult to determine the dust-column density directly.
 We can estimate it based on our models once we know 
$T_{\rm cr}$ which is estimated empirically by the observed colours
(Sect.\,4). Similar empirical approaches based on other observables such as 
atomic and molecular spectra may  be tried, but a problem is how to separate the effects of $T_{\rm cr}$ and $T_{\rm eff}$ on the observables.
Also, beside such empirical estimations,
$ T_{\rm cr} $ may hopefully be determined by
the analysis of the detailed processes of dust growth and its segregation coupled with the dynamical processes of the meteorological scale. 
At present, however, such an {\it ab-initio} approach seems to be more difficult and it is not sure if it provides a more accurate estimation 
of $ T_{\rm cr} $ than the empirical estimation.

Another problem in our empirical approach is that we assumed $ T_{\rm cr} $ to be the same for all the models, but $ T_{\rm cr} $ should certainly depend somewhat on $ T_{\rm eff} $. In fact, our models based on
$ T_{\rm cr} \approx 1800$\,K and extended to $ T_{\rm eff} $ above 2000\,K
showed only minor effect of dust and may fail to explain the observed spectra of  late M dwarfs which are already known to show the effect 
of dust~\cite{t96a}~\cite{j97}. Probably, our estimate of $ T_{\rm cr} \approx 1800$\,K may be valid for L dwarfs from which this result was obtained. We hope, however, that essentially the same approach can be possible to the model photospheres of late M dwarfs with somewhat lower value of $ T_{\rm cr} $. On the contrary, the exact value of $ T_{\rm cr} $ may not be important in the coolest T dwarfs in which the active dust zone is below the observable photosphere, but information on dust will be almost lost from the spectra.

\section{Concluding Remarks}

We believe that we have shown a possibility for the unified model photospheres 
of UCDs including  L and T dwarfs and have finally found an empirical approach 
to consistently treat dust formation and its segregation process in the 
photospheric environment. A natural consequence of our approach
is that dust exists only in the restricted region deep in the photosphere,
and thus a warm dust layer or a cloud is formed in the photosphere.
Our approach is based on a simple assumption that only  small
dust grains can be sustained in the photosphere, and we show that a
self-consistent non-grey model photosphere can be developed without any
other ad-hoc assumption.  
However, once dust appears in the photosphere, it introduces some
inherent difficulties. One problem is that the gas-dust phase change cannot be 
treated properly within the framework of the classical theory of stellar 
atmosphere and in future we should probably learn the recipes of meteorology. 
For example, we said nothing about the fate of 
 the dust grains precipitated below the photosphere, where they may 
evapolate. These large grains may 
give some effects on the photospheric structure as well
as on the observable properties of dusty objects.
Also, unlike atoms and molecules that show well defined spectra, dust shows no 
identifiable spectrum especially in the case of UCDs and a formidable problem 
is how to know the dust-column density in the observable photosphere when we 
apply our models to the fully quantitative analysis of observed spectra of
UCDs.
It is desirable to apply model photospheres with these limitations in mind, 
even though model photospheres can be well useful as a guide for 
interpretation and analysis of the observed data.

\vspace{07mm}

\hspace{-05mm}{\large {\bf Acknowledgements }}
\vspace{03mm}

\hspace{-05mm}I thank Tadashi Nakajima for helpful discussion throughout 
this work 
and Hugh Jones for careful reading of the text with useful comments.
I also thank Tom Geballe, Hugh Jones and Ben Oppenheimer
 for making available their spectra in digital form.

\end{document}